\documentclass[aps,preprint,prl,superscriptaddress]{revtex4}

\usepackage{graphicx}
\usepackage{amsmath}
\usepackage{amssymb}
\usepackage{units}

\usepackage{setspace}

\renewcommand{\vec}[1]{\boldsymbol{#1}}

\begin{document}

\title{Snapshots of non-equilibrium Dirac carrier distributions in graphene}

\author{Isabella Gierz}
\email{Isabella.Gierz@mpsd.cfel.de}
\affiliation{Max Planck Institute for the Structure and Dynamics of Matter, Hamburg, Germany}
\author{Jesse C. Petersen}
\affiliation{Department of Physics, Clarendon Laboratory, University of Oxford, Oxford, United Kingdom}
\author{Matteo Mitrano}
\affiliation{Max Planck Institute for the Structure and Dynamics of Matter, Hamburg, Germany}
\author{Cephise Cacho}
\author{Edmond Turcu}
\author{Emma Springate}
\affiliation{Central Laser Facility, STFC Rutherford Appleton Laboratory, Harwell, United Kingdom}
\author{Alexander St\"ohr}
\author{Axel K\"ohler}
\author{Ulrich Starke}
\affiliation{Max Planck Institute for Solid State Research, Stuttgart, Germany}
\author{Andrea Cavalleri}
\email{Andrea.Cavalleri@mpsd.cfel.de}
\affiliation{Max Planck Institute for the Structure and Dynamics of Matter, Hamburg, Germany}
\affiliation{Department of Physics, Clarendon Laboratory, University of Oxford, Oxford, United Kingdom}

\date{\today}

\maketitle

{\bf The optical properties of graphene are made unique by the linear band structure and the vanishing density of states at the Dirac point \cite{Nair_2008,Bao_2009,Sun1_2010,Ryzhii_2007,Li_2012,Winzer2_2012,Winzer_2010,Winzer_2012,Bonaccorso_2010,Gabor_2011}. It has been proposed that even in the absence of a semiconducting bandgap, a relaxation bottleneck at the Dirac point may allow for population inversion and lasing at arbitrarily long wavelengths \cite{Ryzhii_2007,Li_2012,Winzer2_2012}. Furthermore, efficient carrier multiplication by impact ionization has been discussed in the context of light harvesting applications \cite{Winzer_2010,Winzer_2012}. However, all these effects are difficult to test quantitatively by measuring the transient optical properties alone, as these only indirectly reflect the energy and momentum dependent carrier distributions. Here, we use time- and angle-resolved photoemission spectroscopy with femtosecond extreme ultra-violet (EUV) pulses at 31.5\,eV photon energy to directly probe the non-equilibrium response of Dirac electrons near the $\overline{\text{K}}$-point of the Brillouin zone. In lightly hole-doped epitaxial graphene samples \cite{Riedl_2009,Forti_2011}, we explore excitation in the mid- and near-infrared, both below and above the minimum photon energy for direct interband transitions. While excitation in the mid-infrared results only in heating of the equilibrium carrier distribution, interband excitations give rise to population inversion, suggesting that terahertz lasing may be possible. However, in neither excitation regime do we find indication for carrier multiplication, questioning the applicability of graphene for light harvesting. Time-resolved photoemission spectroscopy in the EUV emerges as the technique of choice to assess the suitability of new materials for optoelectronics, providing quantitatively accurate measurements of non-equilibrium carriers at all energies and wavevectors.}


The non-equilibrium dynamics of Dirac electrons in graphene have so far been primarily investigated with time-resolved optical techniques \cite{Kampfrath_2005,George_2008,Breusing_2009,Wang_2010,Lui_2010,Breusing_2011,Li_2012}, yielding precise estimates of the relaxation rates after photoexcitation. However, as optical probing only indirectly measures energy- and momentum-dependent carrier distributions, controversies have arisen over the underlying physics. Crucially, it has not yet been established if the distribution of hot carriers at early times should be described by a single \cite{Lui_2010}, or by two distinct Fermi-Dirac (FD) distributions with different chemical potentials for valence and conduction band \cite{Breusing_2011,Breusing_2009,Li_2012,Gilbertson_2012}. Clarification of this issue is essential for assessing the potential of graphene for optoelectronics as it defines whether charge carriers in graphene exhibit a metallic or a semiconducting response to photoexcitation. Furthermore, direct experimental proof of the existence of carrier multiplication \cite{Winzer_2010,Winzer_2012} is lacking. 
Here, we resolve these questions by directly mapping the transient occupation of electronic states as a function of momentum and energy with photoemission spectroscopy for different time delays after photoexcitation. 


\begin{figure}
	\center
  \includegraphics[width = 0.7\columnwidth]{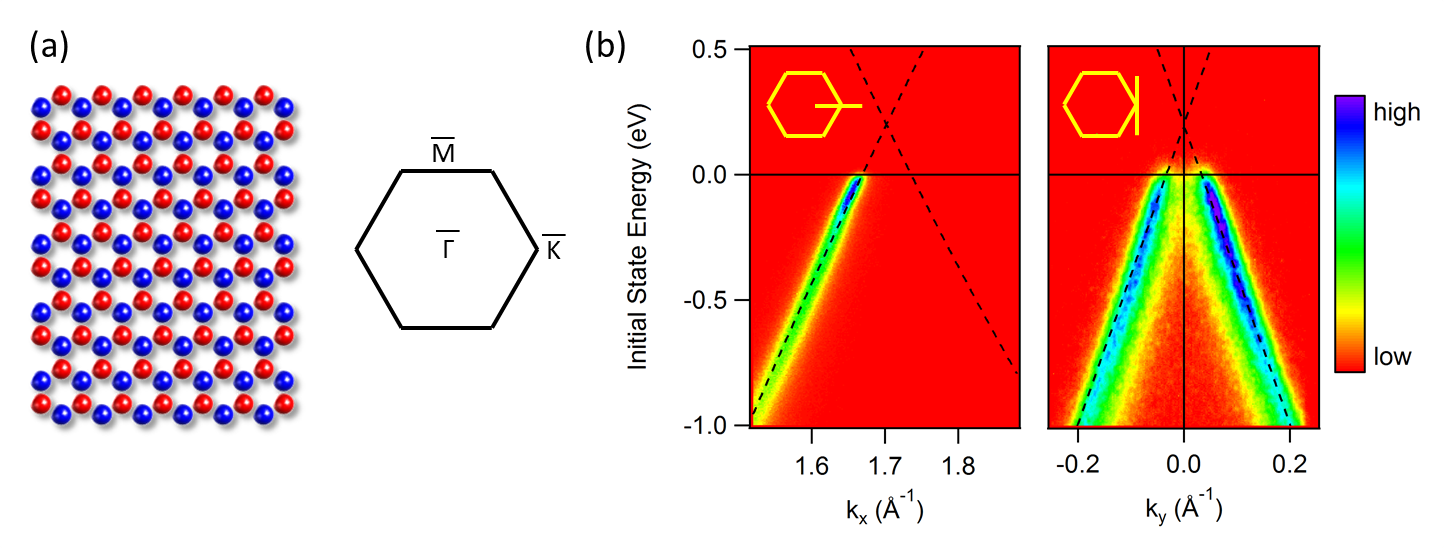}
  \caption{(a) Honeycomb lattice with hexagonal Brillouin zone. The two triangular sublattices are shown in red and blue, respectively. (b) Equilibrium band structure measured with linearly-polarized synchrotron radiation ($\hbar\omega=30$\,eV, left) and He II radiation ($\hbar\omega=41$\,eV, right). The color scale is linear with violet (red) corresponding to high (low) photocurrent. The inset shows the cut through the Brillouin zone along which the photoemission data has been measured. Along the $\overline{\Gamma\text{K}}$-direction the photocurrent is suppressed for one of the two $\pi$-bands \cite{Shirley_1995}. The graphene layer is slightly hole-doped so that the chemical potential $\mu_e$ is $\sim$200\,meV below the Dirac point. Dashed black lines represent the dispersion obtained from a tight-binding model \cite{Bostwick_2007}.}
  \label{figure1}
\end{figure}

The honeycomb lattice of graphene, together with a sketch of the two-dimensional Brillouin zone is shown in Fig. \ref{figure1}a. We studied a quasi-freestanding epitaxial graphene monolayer grown on SiC(0001) \cite{Riedl_2009,Forti_2011}, which was first characterized by static angle-resolved photoemission spectroscopy (ARPES). Figure \ref{figure1}b shows photoelectron energy and momentum maps along and perpendicular to the $\overline{\Gamma\text{K}}$-direction, evidencing the well-known conical band structure \cite{Bostwick_2007,Bostwick_2010}. For measurements perpendicular to the $\overline{\Gamma\text{K}}$-direction, extrapolation of the bands based on a tight-binding model (dashed lines) \cite{Bostwick_2007} shows that the graphene layer is hole-doped, with the Dirac point $\sim$200\,meV above the static chemical potential $\mu_e$. For measurements along $\overline{\Gamma\text{K}}$, only one of the two $\pi$-bands is visible due to photoemission matrix element effects \cite{Shirley_1995}. 

The non-equilibrium response of this sample was studied for excitation below and above the minimum photon energy for interband excitations ($2\mu_e=400$\,meV). In this way, both {\it metallic} ($\hbar\omega_{\text{pump}}=300$\,meV) and {\it semiconducting} ($\hbar\omega_{\text{pump}}=950$\,meV) optical properties were explored. To probe the dynamics of the $\pi$-bands at the $\overline{\text{K}}$-point of the hexagonal Brillouin zone ($k_{||}=1.7$\AA$^{-1}$, requiring $\hbar\omega>16$\,eV), 30\,femtosecond (fs) EUV pulses from a High Harmonic Generation (HHG) source were used \cite{Mathias_2007,Rohwer_2011,Petersen_2011}. A time preserving monochromator selected a single harmonic at $\hbar\omega_{\text{probe}}=31.5$\,eV with energy resolution of 130\,meV \cite{Frassetto_2011}. 

\begin{figure}
	\center
  \includegraphics[width = 0.6\columnwidth]{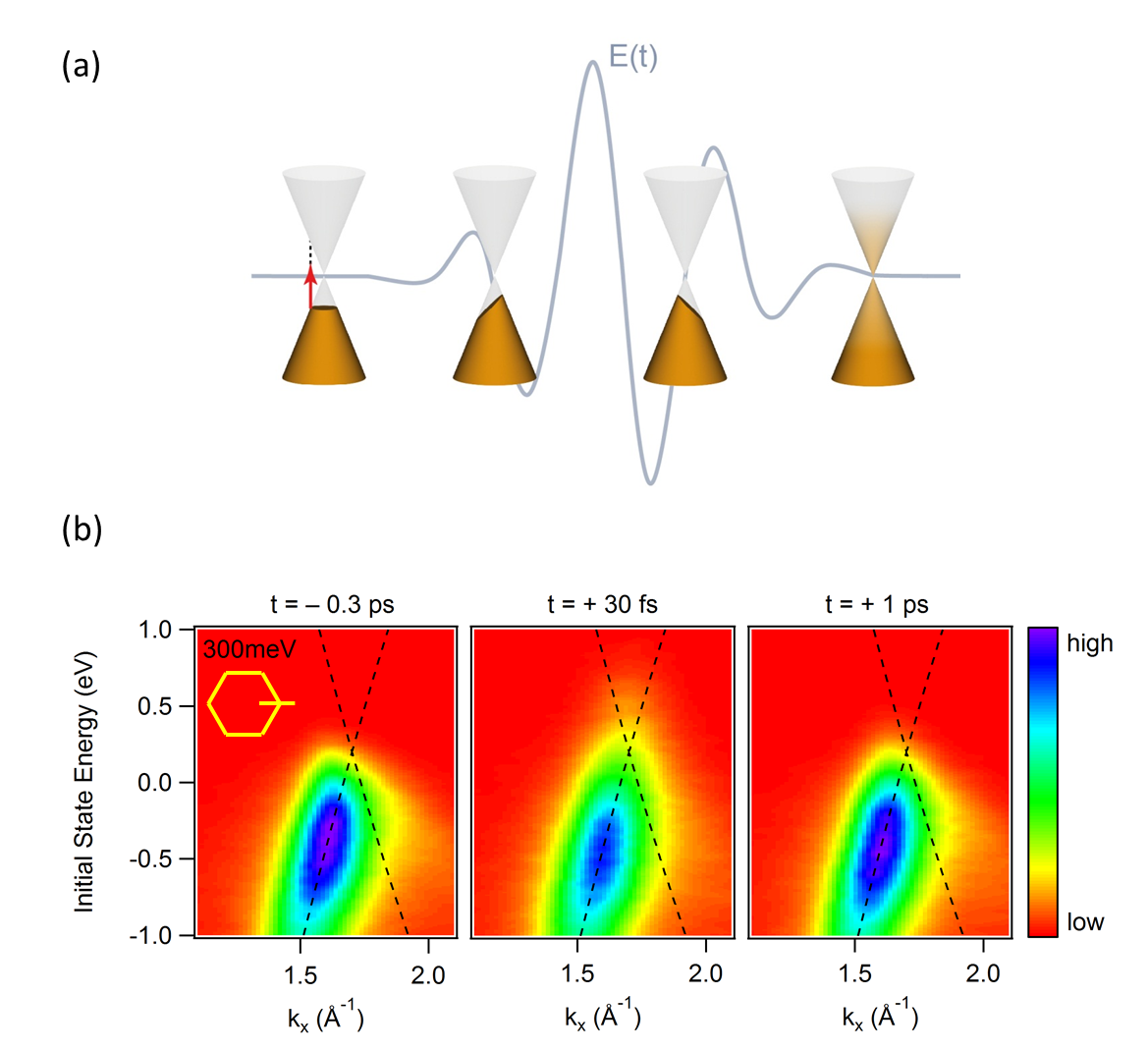}
  \caption{Free carrier absorption regime for excitation at $\hbar\omega_{\text{pump}}=300$\,meV: (a) Cartoon of the excitation mechanism. For $\hbar\omega_{\text{pump}} < 2|\mu_e|$ direct interband transitions are impossible. The Fermi liquid sloshes around the Dirac cone following the electric field $E(t)$ of the laser pulse (light grey line) like water in a shaking glass. Electron-electron and electron-phonon scattering events subsequently establish a thermal Fermi-Dirac distribution. Within the time resolution of the present investigation the coherent dynamics at earliest times could not be resolved. (b) Snapshots of the band structure (smoothed) close to the $\overline{\text{K}}$-point for different pump-probe delays: negative time delays (left), close to zero delay ($+30$\,fs, middle), $t=1$\,ps (right). The sample is excited at $\hbar\omega_{\text{pump}}=300$\,meV with a fluence of $F=0.8$\,mJ/cm$^2$. Photoelectrons are ejected using photons of energy $\hbar\omega_{\text{probe}}=31.5$\,eV. The measurements were done along the $\overline{\Gamma\text{K}}$-direction (inset), where the photocurrent is suppressed for one of the two $\pi$-bands \cite{Shirley_1995}. Dashed black lines represent the dispersion obtained from a tight-binding model \cite{Bostwick_2007}.}
  \label{fig_snapshots_300meV}
\end{figure}

Figure \ref{fig_snapshots_300meV}a shows a cartoon of the expected excitation mechanism for photon energies $\hbar\omega_{\text{pump}} < 2|\mu_e|$, for which only free carrier absorption is possible. For an electromagnetic pulse with vector potential $\vec{A}(t)=\int \vec{E} dt$, the Dirac electrons are accelerated and decelerated according to the interaction Hamiltonian $H=\frac{1}{2m^*}(\vec{p}-q\vec{A})^2$. Scattering events with other electrons or with the lattice cause rapid thermalization into a hot carrier distribution. 

A set of experimental snapshots of the Dirac cone is plotted in Fig. \ref{fig_snapshots_300meV}b for different pump-probe delays after excitation at $\hbar\omega_{\text{pump}}=300$\,meV with a fluence of $F=0.8$\,mJ/cm$^2$. Within the time resolution of our measurement we did not resolve non-thermal energy distributions for the Dirac electrons, indicating very efficient thermalization. Rather, we observed instantaneous broadening of the carrier distribution through the Dirac point (Fig. \ref{fig_snapshots_300meV}b middle panel) and subsequent relaxation within $\sim$1 picosecond (ps, Fig. \ref{fig_snapshots_300meV}b right panel).

\begin{figure}
	\center
  \includegraphics[width = 1\columnwidth]{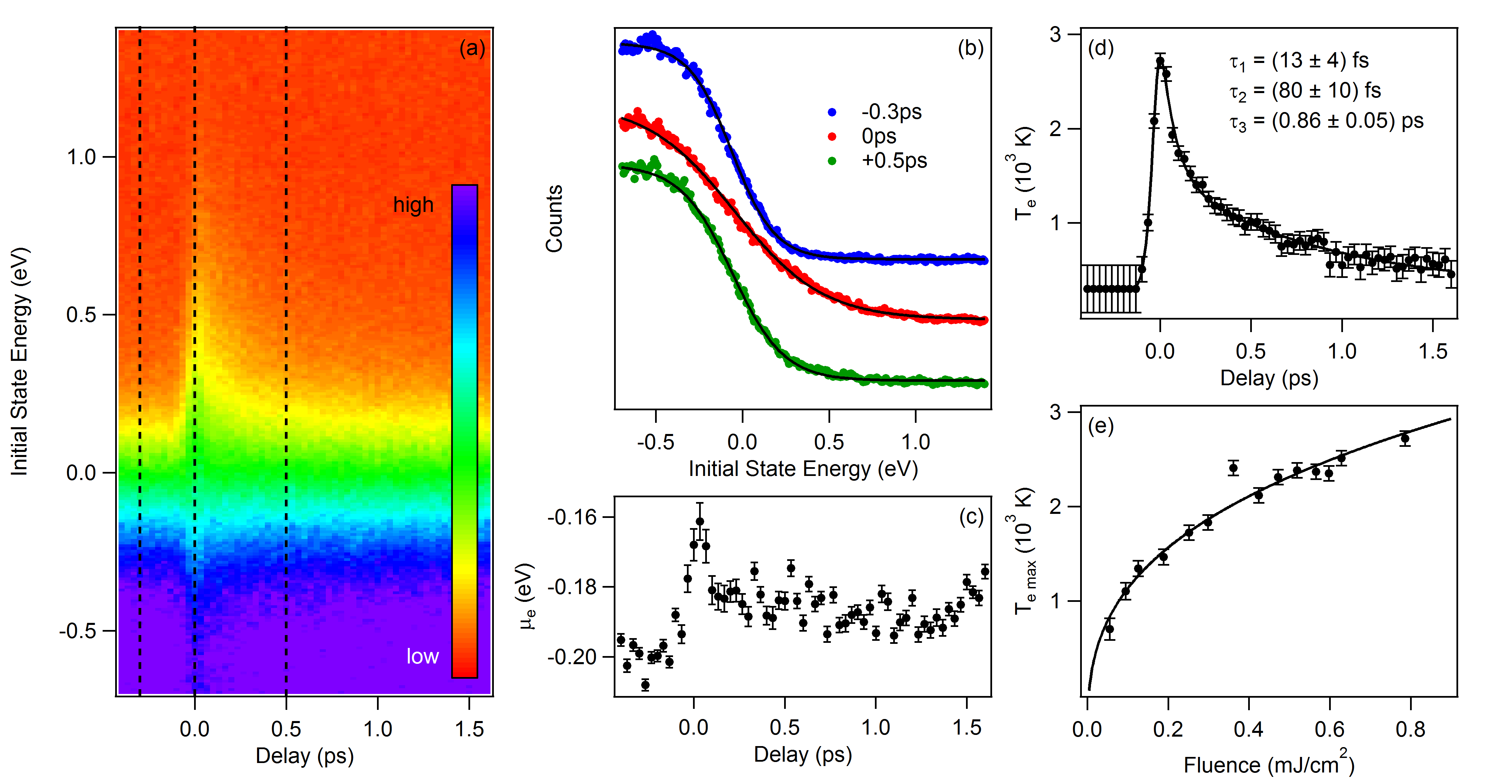}
  \caption{Metallic response for excitation at $\hbar\omega_{\text{pump}}=300$\,meV and $F=0.8$\,mJ/cm$^2$: (a) Energy-distribution curves (EDCs) integrated over the momentum range displayed in Fig. \ref{fig_snapshots_300meV}b as a function of delay in a two-dimensional color plot. (b) Selected EDCs extracted along the dashed lines in panel (a) together with Fermi-Dirac distribution fits that are shown as solid black lines. Chemical potential (c) and electronic temperature (d) as a function of delay. The solid black line in (d) is a fit including an error-function for the rise time and three exponential decays: $\tau_1=(13\pm4)$\,fs, $\tau_2=(80\pm10)$\,fs and $\tau_3=(0.86\pm0.05)$\,ps, consistent with the results of all-optical measurements. (e) The fluence dependence of the maximum electronic temperature follows the $F^{1/3}$ behavior (black line) expected for free carrier absorption.}
  \label{fig_metallic_regime}
\end{figure}

A quantitative analysis of these dynamics is shown in Fig. \ref{fig_metallic_regime}. In panel (a) we display energy distribution curves (EDCs) that have been integrated at each time delay over the momentum range depicted in Fig. \ref{fig_snapshots_300meV}b. Zero delay corresponds to the maximum pump-probe signal. The energy dependence of the EDCs is well fitted at all times by the FD distribution $f(E,\mu_e,T_e)=\left(\exp\left(\frac{E-\mu_e}{k_BT_e}\right)+1\right)^{-1}$, where $E$ is the energy, $\mu_e$ is the chemical potential, $k_B$ is the Boltzmann constant, and $T_e$ is the electronic temperature. The broadening of the Fermi edge immediately after absorption of the pump pulse reflects an increase in electronic temperature $T_e$ that subsequently cools down to room temperature within $\sim$1\,ps. Figure \ref{fig_metallic_regime}b displays EDCs for selected time delays [-0.3 ps (blue), 0 ps (red) and +0.5 ps (green)] together with FD fits (black lines). Fit results for the chemical potential $\mu_e$ and the electronic temperature $T_e$ are shown in Figs. \ref{fig_metallic_regime}c and d, respectively. 

The chemical potential initially increases by approximately 40\,meV, recovering as the carriers cool. The initial increase and fast decay of the chemical potential is well understood as a consequence of the changes in electronic temperature (Fig. \ref{fig_metallic_regime}d), as $\Delta\mu=\frac{\pi^2}{6}(k_B T_e)^2/E_F$ \cite{Ashcroft_Mermin}, where $E_F$ is the Fermi energy. 

The time-dependent electronic temperature $T_e(t)$ is shown in Fig. \ref{fig_metallic_regime}d. $T_e$ is observed to instantly rise to nearly 3000\,K, before decaying with a triple exponential law (black line in Fig. \ref{fig_metallic_regime}d), in excellent agreement with previous all-optical measurements. These measurements have attributed the decay channels to electron-electron scattering ($\tau_1$) \cite{Breusing_2009,Breusing_2011}, emission of optical phonons ($\tau_2$) \cite{Kampfrath_2005,Butscher_2007,George_2008,Lui_2010}, and further cooling through the decay of optical into acoustic phonons ($\tau_3$) \cite{Bonini_2007,Yan_2009,Wang_2010,Sun_2010,Kang_2010}.  More recently, the third relaxation time $\tau_3$ has been ascribed to the direct coupling between electrons and acoustic phonons in the presence of lattice defects \cite{Song_2012,Graham_2013}. Our data is consistent with both interpretations of $\tau_3$. A validation of the decay law would require temperature-dependent measurements that are beyond the scope of his paper.
 
The fluence dependence of the peak electronic temperature, which further substantiates the physical picture discussed above, is shown in Fig. \ref{fig_metallic_regime}e. The increase in kinetic energy of the Dirac carriers can be obtained from $\Delta E(T_e)=\int_{\mu_e}^{\infty}DOS(E)f(E,\mu_e,T_e)E dE$, where $DOS(E)\propto|E|$ is the linear density of states of graphene. For simplicity, we choose $\mu_e(T_e)=0$ and obtain $\Delta E\propto T_e^3$, where $\Delta E$ is equivalent to the absorbed fluence $F$ in mJ/cm$^2$. The experimental fluence dependence of $T_e$, displayed in Fig. \ref{fig_metallic_regime}e, follows the expected $F^{1/3}$ behavior (black line).

\begin{figure}
	\center
  \includegraphics[width = 0.6\columnwidth]{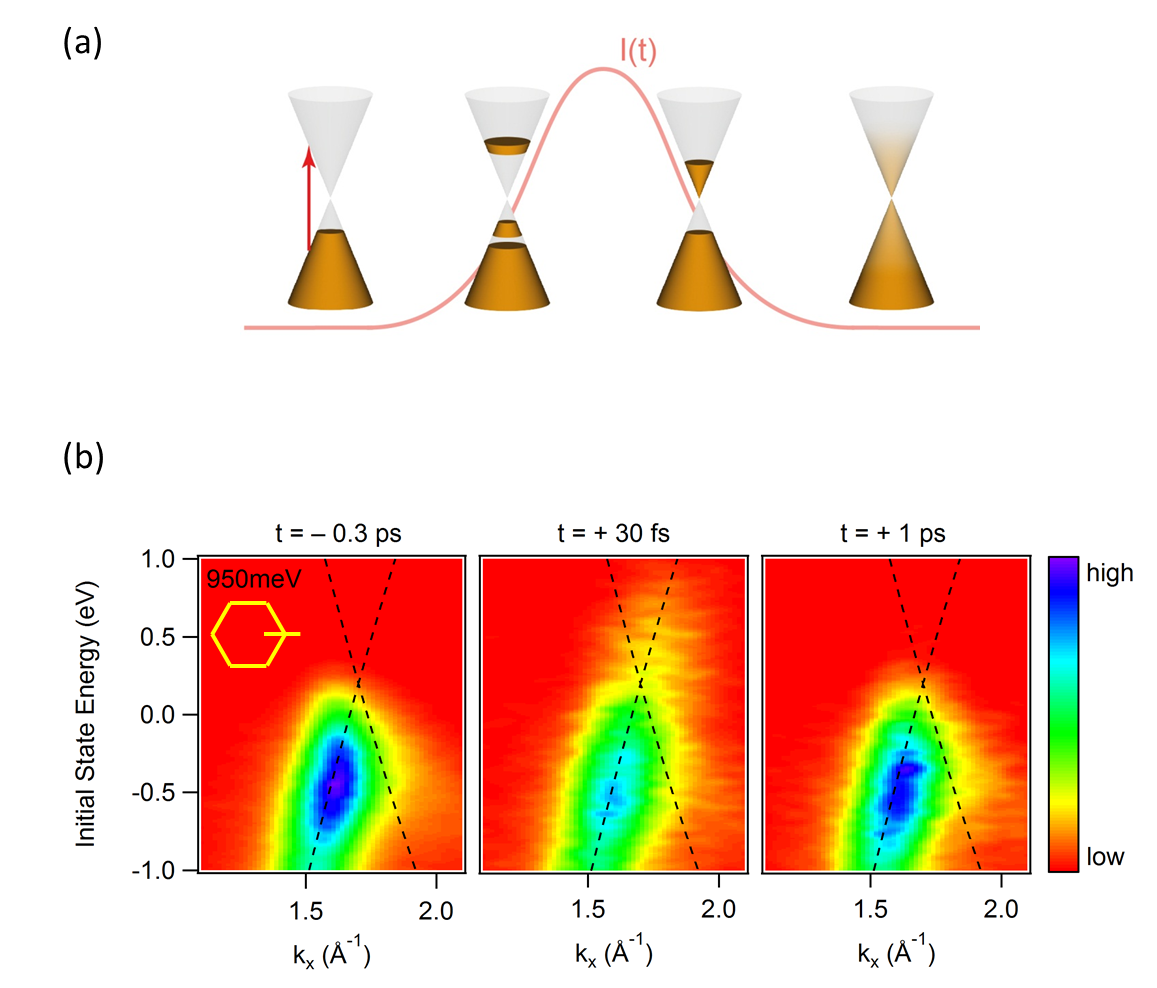}
  \caption{Direct interband transition regime for excitation at $\hbar\omega_{\text{pump}}=950$\,meV: (a) Cartoon of the excitation mechanism. The red line represents the intensity envelope of the pulse. If $\hbar\omega_{\text{pump}}>|2\mu_e|$, the graphene layer is excited via direct interband transitions that leave the system with a highly non-thermal electron distribution that cannot be observed within the temporal resolution of the present experiment. Within $\sim$10\,femtoseconds the electrons thermalize and form separate Fermi-Dirac distributions for the valence and conduction band. After some hundreds of femtoseconds the separate chemical potentials merge, resulting in a single Fermi-Dirac distribution for valence and conduction band. (b) Snapshots of the band structure (smoothed) for excitation at $\hbar\omega_{\text{pump}}=950$\,meV with a fluence of $F=4.6$\,mJ/cm$^2$ and a probe photon energy of $\hbar\omega_{\text{probe}}=31.5$\,eV. The measurements were done along the $\overline{\Gamma\text{K}}$-direction (inset). Dashed black lines represent the dispersion obtained from a tight-binding model \cite{Bostwick_2007}.}
  \label{fig_snapshots_950meV}
\end{figure}

\begin{figure}
	\center
  \includegraphics[width = 1\columnwidth]{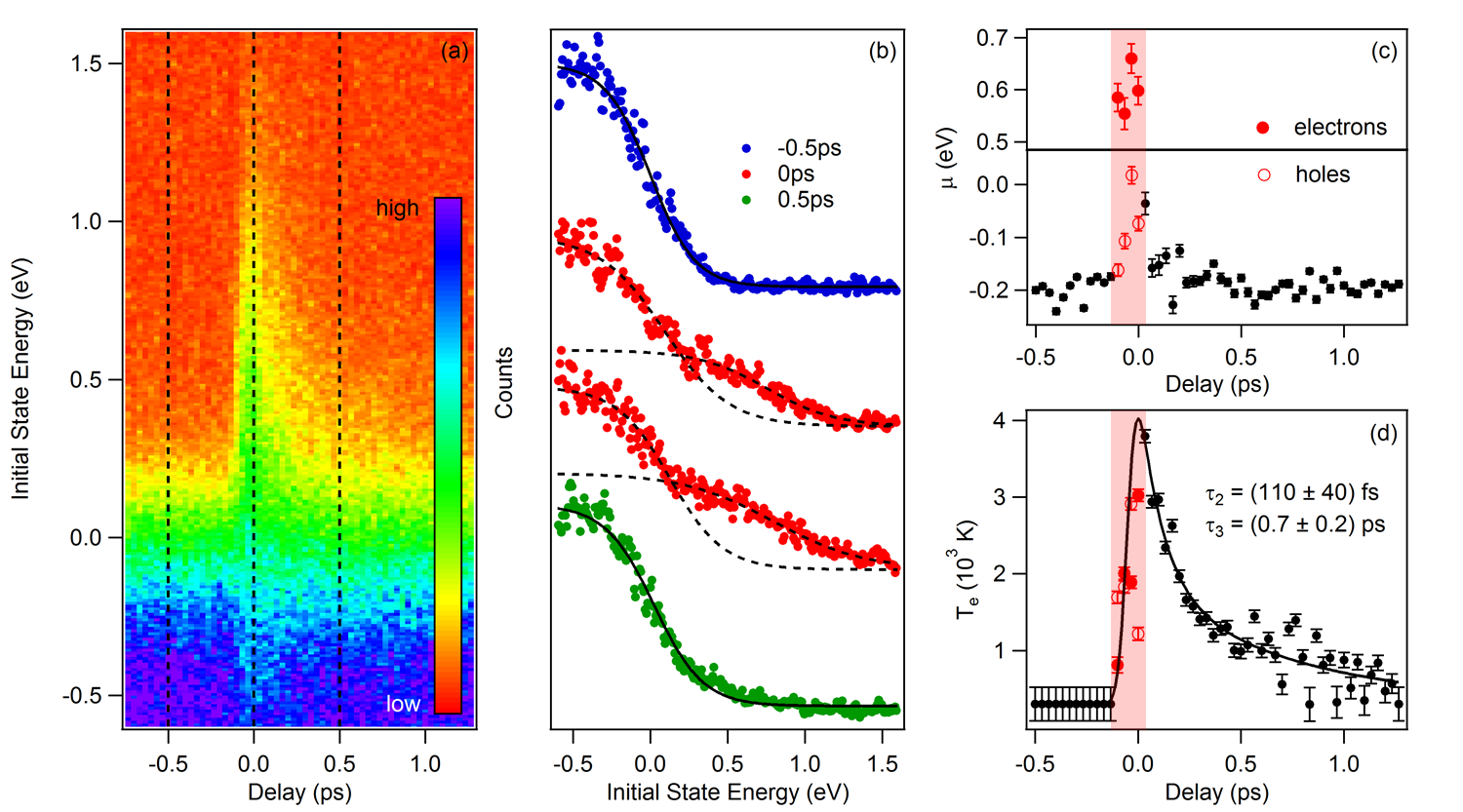}
  \caption{Population inversion for excitation at $\hbar\omega_{\text{pump}}=950$\,meV and $F=4.6$\,mJ/cm$^2$: (a) Momentum-integrated energy distribution curves (EDCs) as a function of delay. (b) EDCs for three selected delays extracted along the dashed lines in (a) together with Fermi-Dirac distribution fits. The EDCs close to time zero are best fitted with two distinct Fermi-Dirac distributions for electrons and holes (dashed black lines). Chemical potential (c) and electronic temperature (d) as a function of delay. The red-shaded area indicates the time interval over which two distinct Fermi-Dirac distributions persist. The solid black line in panel (d) represents a fit of the black data points with a fixed rise time of 50\,fs and a double exponential decay. The decay times are indicated in the figure.}
  \label{fig_population_inversion}
\end{figure}

We next report results for direct interband excitation at $\hbar\omega_{\text{pump}}=950$\,meV in Figs. \ref{fig_snapshots_950meV} and \ref{fig_population_inversion}. Figure \ref{fig_snapshots_950meV}a displays a schematic of the expected interband excitation mechanism, which, in contrast to the free carrier absorption discussed above, is driven by the intensity envelope $I(t)$ of the laser pulse (red line) rather than the electric field. During interband excitation, a non-thermal distribution of excited electron-hole pairs is generated at the earliest time delays, relaxing into a quasi-equilibrium state with a temperature higher than that of the lattice. Importantly, depending on the relative strength of interband versus intraband scattering, a single metallic distribution like the one of Fig. \ref{fig_metallic_regime}b, or two distinct distributions for valence and conduction band are attained. Only the latter scenario would make light amplification possible.
 
In Fig. \ref{fig_snapshots_950meV}b we present snapshots of the Dirac carrier distributions for different time delays after excitation at $\hbar\omega_{\text{pump}}=950$\,meV and a fluence of $F=4.6$\,mJ/cm$^2$. The excitation fluence was chosen according to experimental estimates of the threshold for optical gain in Ref. \cite{Li_2012}. 

Figures \ref{fig_population_inversion}a and \ref{fig_population_inversion}b report momentum-integrated EDCs as a function of pump-probe delay. In contrast to Fig. \ref{fig_metallic_regime}b, the carrier distributions measured immediately after excitation are best fitted by two separate FD distributions for valence and conduction band, indicating population inversion \cite{Li_2012}. The fit results for the chemical potential $\mu_e$ and the electronic temperature $T_e$ are displayed in Figs. \ref{fig_population_inversion}c and \ref{fig_population_inversion}d, respectively. The two distributions merge within $\sim$130\,fs, and a single FD distribution is attained, promoted by scattering of charge carriers across the Dirac point. As amplification is only possible for the given time window, our measurement sets a quantitative boundary to be used as a benchmark for any laser design. 

\begin{figure}
	\center
  \includegraphics[width = 0.7\columnwidth]{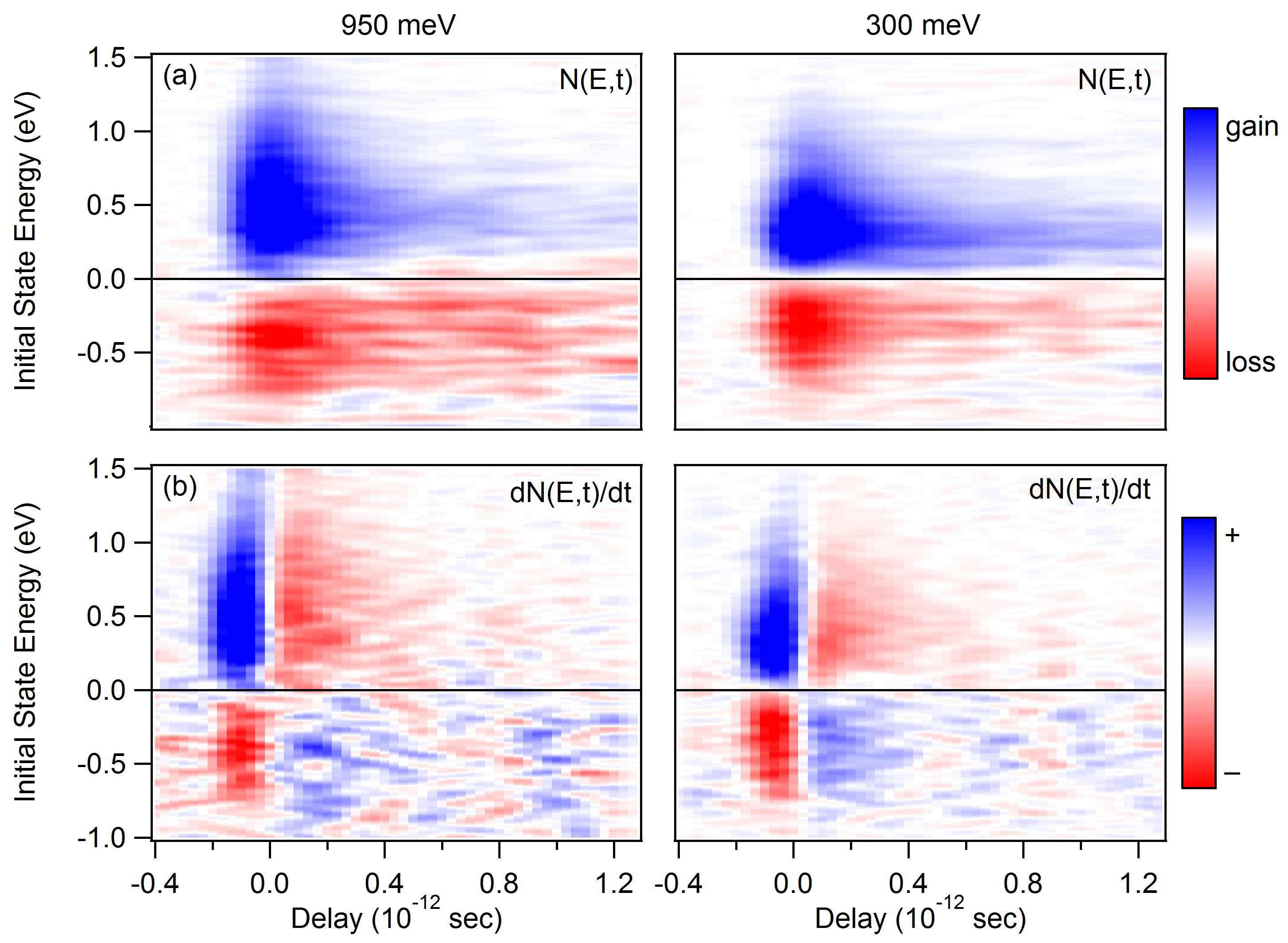}
  \caption{No indication of carrier multiplication for direct interband transitions (left column) and free carrier absorption (right column):  The pump-probe signal in panel (a) is given by the difference between the photocurrent shown in Figs. \ref{fig_metallic_regime}a and \ref{fig_population_inversion}a, respectively, and the photocurrent at negative time delays. Panel (b) shows the time derivative of the pump-probe signal, which is equivalent to the electron-hole pair generation/recombination rate. Electron-hole pairs are generated only for a short time interval in the presence of the pump pulse. The data has been smoothed.}
  \label{fig_carrier_multiplication}
\end{figure}

Our measurements also provide a direct assessment of the possibility of carrier multiplication in graphene \cite{Winzer_2010,Winzer_2012}. Figure \ref{fig_carrier_multiplication} displays the time-dependent evolution of the pump-probe signal for the two cases of free carrier absorption (right panel) and interband excitation (left panel). Specifically, Fig. \ref{fig_carrier_multiplication}a shows the pump-induced changes of the photocurrent (directly proportional to the number of electrons $N$ at energy $E$) as a function of energy and time delay. Blue (red) corresponds to a gain (loss) in photocurrent with respect to negative time delays. In Fig. \ref{fig_carrier_multiplication}b, the time derivative of the pump-probe signal of panel (a) is displayed, directly revealing the electron-hole pair generation/recombination rates. The generation/recombination rate $dN(E > 0,t)/dt$ is positive only for a few tens of femtoseconds. We conclude that electron-hole pairs are generated only while the pump pulse is present, whereas no spontaneous carrier multiplication is found.

\begin{figure}
	\center
  \includegraphics[width = 0.5\columnwidth]{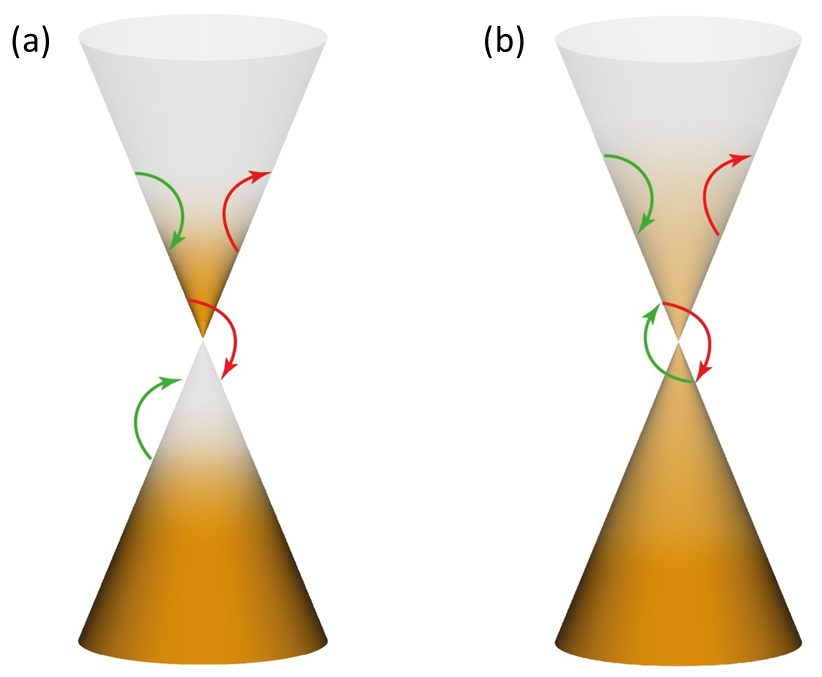}
  \caption{Competing relaxation channels in graphene: For a population-inverted state with separate chemical potentials for valence and conduction band (a) impact ionization (green arrows) can only create additional electron-hole pairs if $k_B T_e \sim |\mu_e|+|\mu_h|$, which requires temperatures above $\sim$8000\,K, two times larger than the measured peak temperature. Auger heating (red arrows) dominates and leads to the formation of a single hot electronic distribution (b). In this case impact ionization and Auger heating are equally likely with zero net effect, and the electronic system can only cool down via phonon emission.}
  \label{figure7}
\end{figure}

The absence of carrier multiplication is well understood when analyzing the fits of Fig. \ref{fig_population_inversion}. Indeed, in the presence of two distinct FD distributions, impact ionization \cite{Winzer_2010,Winzer_2012} can only create additional electron-hole pairs if the temperature of the Dirac carriers $k_B T_e$ exceeds the sum of the electron and hole chemical potentials $|\mu_e|+|\mu_h|$ (Fig. \ref{figure7}a). This would require temperatures above $\sim$8000\,K, significantly higher than the peak temperature of $\sim$4000\,K estimated in Fig. \ref{fig_population_inversion}. Instead, the inverse scattering process (Auger heating) dominates, rapidly merging the separate chemical potentials. For a single FD distribution (Fig. \ref{figure7}b) attained $\sim$130\,fs after direct interband transitions or at all times for free carrier absorption, the probabilities for impact ionization and Auger heating are similar, again preventing carrier multiplication. 
The absence of carrier multiplication and the short lifetime of electron-hole pairs discussed above raises serious doubts about the suitability of graphene for efficient light harvesting for the present excitation regime. One may speculate that more favorable conditions for carrier multiplication may be met for negligible doping of the graphene layer, smaller pump fluences and higher excitation energies \cite{Winzer_2012}.

In summary, we have used time- and angle-resolved photoemission in the EUV as a tool to critically assess the potential of graphene for optoelectronic applications. Two key contributions are presented.

Firstly, we show that population inversion occurs immediately after interband excitation, substantiating the view of graphene as a zero bandgap semiconductor. Previous time-resolved optical experiments had evidenced negative optical conductivity (optical gain) \cite{Li_2012}. However, these measurements only probed at higher photon energies and could not detect the dynamics close to the Dirac point. The present experiment allows for a direct measurement down to arbitrarily small energies and with a temporal resolution that would not be possible for terahertz optical probes. Secondly, by directly measuring the band occupancy $N(E,t)$ at all energies and times, we have made a quantitative evaluation of the relevance of carrier multiplication. 

Direct mapping of the non-equilibrium carrier distribution and primary acceleration/scattering events with few femtosecond or even attosecond resolution \cite{Cavalieri_2007} is clearly a highly interesting area of future research. Combining ultrafast photoemission techniques with nanoscale spatial resolution, for example in Photoemission Electron Microscopy (PEEM), would make it possible to investigate nanoscale carrier dynamics as well as real device transport.

\section*{Methods}

Details about the sample growth are reported in Ref. \cite{Riedl_2009,Forti_2011}. The SiC(0001) substrate was etched in Hydrogen atmosphere to produce atomically flat terraces. In a second step, the substrate was annealed in inert Argon gas to grow a homogeneous carbon monolayer that was subsequently decoupled from the substrate by Hydrogen intercalation, resulting in a completely sp$^2$-hybridized graphene layer. After growth, the samples have been transported in air and inserted into an ultra-high vacuum chamber, where the samples were annealed at about 500$^{\circ}$C to remove adsorbates, and characterized with static ARPES. Measurements at the $\overline{\text{K}}$-point perpendicular to the $\overline{\Gamma\text{K}}$-direction were performed at the Max Planck Institute for Solid State Research in Stuttgart (Germany) with Helium II radiation at $\hbar\omega=41$\,eV. The static ARPES measurement along the $\overline{\Gamma\text{K}}$-direction was done at the Swiss Light Source (SLS) of the Paul Scherrer Institute (Villigen, Switzerland) with a photon energy of $\hbar\omega=30$\,eV and linearly polarized light. Measurements at the SLS were performed on a different sample \cite{Forti_2011} (but grown after the same recipe) than the other static and time-resolved experiments in the manuscript.

Time-resolved ARPES measurements were performed at the materials science end station at Artemis (Central Laser Facility, Didcot, United Kingdom), using synchronized near-infrared (NIR) and EUV photons as pump and probe pulses, respectively. The facility is equipped with a 1\,kHz Ti:Sapphire laser system that delivers pulses with a central wavelength of 790\,nm and a nominal pulse duration of 30 femtoseconds. NIR photons are generated via optical parametric amplification (OPA, 950\,meV) and subsequent difference frequency generation (DFG, 300\,meV). EUV photons in the range from 20 to 40\,eV are obtained via high harmonics generation in an Argon gas jet. For the present experiment a photon energy of $\hbar\omega_{\text{probe}}=31.5$\,eV has been chosen, using a time preserving monochromator to maintain a short pulse, enabling access to electrons emitted at the $\overline{\text{K}}$-point of the two-dimensional graphene Brillouin zone.

All measurements in this paper were performed at room temperature, except the static dispersion along the $\overline{\Gamma\text{K}}$-direction shown in Fig. 1(b) which was acquired at a sample temperature of 70\,K.

\section*{Acknowledgments}

We thank Martin Eckstein and Franz K\"artner for many fruitful discussions and J\"org Harms for drawing Figs. \ref{fig_snapshots_300meV}a and \ref{fig_snapshots_950meV}a. Stiven Forti, Camilla Coletti and Konstantin V. Emtsev helped with the static ARPES measurements at the SLS, partially supported by the German Research Foundation (DFG) within the priority program `graphene' SPP 1459 (Sta 315/8-1). Phil Rice and Natercia Rodrigues are acknowledged for technical support during the Artemis beamtime.

\section*{Author Contributions}

I.\,G., J.\,C.\,P., M.\,M. and C.\,C. performed the time-resolved experiments. E.\,S. managed the laboratory and E.\,T. ran the laser system; both E.\,T. and E.\,S. provided technical support during the beamtime. A.\,S., A.\,K. and U.\,S. grew and characterized the samples. I.\,G. analyzed the data. I.\,G. and A.\,C. interpreted the results and wrote the manuscript.

\section*{Author Information}

The authors declare no competing financial interests. Correspondence and requests for materials should be addressed to I.\,G. (isabella.gierz@mpsd.cfel.de).

\end{document}